\documentclass[manuscript]{aastex}
\usepackage[fleqn]{amsmath}

\newcommand{\lc}{light curve~}
\newcommand{\lct}{light curve}

\newcommand{\mjup}{\ensuremath{M_{\rm J}}}

\newcommand{\ik}{{\it Kepler~}}
\newcommand{\ikt}{{\it Kepler}}

\newcommand{\figr}[1]{Fig.~\ref{fig:#1}}
\newcommand{\secr}[1]{\mbox{\S\ \ref{sec:#1}}}
\newcommand{\eqr}[1]{Eq.~\ref{eq:#1}}

\newcommand{\tabr}[1]{\mbox{Table~\ref{tab:#1}}}

\received{August 16, 2011}
\accepted{October 4, 2011}

\shorttitle{KOI-13 Orbital Photometry}

\shortauthors{Shporer et al.}

\begin{document}

\title{Detection Of KOI-13.01 Using The Photometric Orbit}

\author{Avi~Shporer\altaffilmark{1, 2}, 
Jon~M.~Jenkins\altaffilmark{3},
Jason~F.~Rowe\altaffilmark{4},
Dwight~T.~Sanderfer\altaffilmark{4},
Shawn~E.~Seader\altaffilmark{3},
Jeffrey~C.~Smith\altaffilmark{3},
Martin~D.~Still\altaffilmark{5},
Susan~E.~Thompson\altaffilmark{3},
Joseph~D.~Twicken\altaffilmark{3},
and William~F.~Welsh\altaffilmark{6}
} 
\altaffiltext{1}{Las Cumbres Observatory Global Telescope Network, 6740 Cortona Drive, Suite 102, Santa Barbara, CA 93117, USA; ashporer@lcogt.net}
\altaffiltext{2}{Department of Physics, Broida Hall, University of California, Santa Barbara, CA 93106, USA}
\altaffiltext{3}{SETI Institute/NASA Ames Research Center, Moffett Field, CA 94035, USA}
\altaffiltext{4}{NASA Ames Research Center, Moffett Field, CA 94035, USA}
\altaffiltext{5}{Bay Area Environmental Research Institute/NASA Ames Research Center, Moffett Field, CA 94035, USA}
\altaffiltext{6}{Astronomy Department, San Diego State University, San Diego, CA 92182, USA}

\begin{abstract}

We use the KOI-13 transiting star-planet system as a test case for the recently developed BEER algorithm (Faigler \& Mazeh 2011), aimed at identifying non-transiting low-mass companions by detecting the photometric variability induced by the companion along its orbit. Such photometric variability is generated by three mechanisms, including the beaming effect, tidal ellipsoidal distortion, and reflection/heating. We use data from three \ik quarters, from the first year of the mission, while ignoring measurements within the transit and occultation, and show that the planet's ephemeris is clearly detected. We fit for the amplitude of each of the three effects and use the beaming effect amplitude to estimate the planet's minimum mass, which results in $M_p\sin i = 9.2  \pm 1.1 $ \mjup\ (assuming the host star parameters derived by Szabo et al.~2011).
Our results show that {\it non-transiting} star-planet systems similar to KOI-13.01 can be detected in \ik data, including a measurement of the orbital ephemeris and the planet's minimum mass. Moreover, we derive a realistic estimate of the amplitudes uncertainties, and use it to show that data obtained during the entire lifetime of the \ik mission, of 3.5 years, will allow detecting non-transiting close-in low-mass companions orbiting bright stars, down to the few Jupiter mass level. Data from the \ik Extended Mission, if funded by NASA, will further improve the detection capabilities.


\end{abstract}


\section{Introduction}
\label{intro}

Traditionally, before the age of continuous high-precision space-based photometry, the investigation of low-mass transiting companions used spectroscopy primarily to study the system's orbit via the spectroscopic Doppler effect, and photometry primarily to study the system's geometric characteristics through measurements of eclipse (transit and occultation) light curve. However, it has been shown that space-based photometry from missions such as \ik \citep{borucki10, koch10} and CoRoT \citep{rouan98, auvergne09} can be used to detect optical variability induced by a low-mass companion along its orbital motion, down to the planetary mass level \citep[e.g.,][]{jenkins03, mazeh10, welsh10}. Hereafter we refer to the measurement of photometric variability correlated with the orbit as {\it orbital photometry}.

Recently it was shown by \cite{faigler11} that analysis of orbital photometry can be used for detecting new {\it non-eclipsing} binary systems, where the secondary (star or planet) can be of low-mass, down to the planet mass range. Their BEER algorithm is aimed at identifying the system's ephemeris, and estimating the secondary minimum mass, $M_2\sin i$, where $M_2$ is the mass of the secondary and $i$ the angle between the orbital angular momentum axis and the line of sight. In addition, \cite{faigler11} presented the detection of two targets showing an orbital photometry signal, which in both cases suggests the presence of a non-eclipsing companion whose minimum mass is close to 70 $\mjup$. We use here the same method, with some modifications and additions, and apply it to the recently discovered transiting planet system KOI-13.01 \citep{rowe11a, rowe11b}, containing a massive planet orbiting a bright A-type host star ($K_p = 9.96$ mag; KIC ID 9941662) every 1.76 days. Since that system is transiting we use it as a test case for the approach of \cite{faigler11}. We use only the out-of-eclipse light curve to show whether {\it non-transiting} star-planet systems similar to KOI-13.01 can be discovered in \ik data. 


We briefly describe the various mechanisms inducing photometric variability along the orbit in \secr{orbphot}. In \secr{dataanal} we present our data analysis where we show that the out-of-eclipse data can be used to clearly detect the ephemeris of KOI-13.01. In our discussion in \secr{dis} we give an estimate of the companion minimum mass (which in this case is close to the true mass), and show that data from the entire 3.5 year \ik mission lifetime is expected to  allow detecting non-transiting close-in star-planet systems down to the Jupiter mass level. Our work is summarized in \secr{sum}.

\section{Orbital Photometry}
\label{sec:orbphot}


As described by \cite{faigler11} photometric variability along the orbit can be induced by three mechanisms: (I) beaming (aka Doppler boosting), (II) tidal ellipsoidal distortion, and (III) reflection of light from the star by the planet. The third mechanism includes the effect of heating of the planetary atmosphere by the star's radiation.

The tidal ellipsoidal and reflection effects are well known from observations of eclipsing binaries, while the beaming effect was observed only recently. The latter was described in \cite{shakura87}, \cite{loeb03}, and \cite{zucker07}, and observed from both space \citep{vankerkwijk10, mazeh10, bloemen11, carter11, faigler11} and the ground (\citealt{shporer10}; see also \citealt{maxted00}).

The equations describing the amplitudes, in relative flux, of each of the three effects are given by \cite{faigler11}. We list them here for completeness, for a Jupiter mass companion:
\begin{align}
\label{eq:abeam}
& \hspace{-10mm} A_{\rm beam} = \alpha_{\rm beam} 4 \frac{K_{RV}}{c} = 2.7\ \alpha_{\rm beam} \left(\frac{M_s}{M_{\sun}}\right)^{-2/3} \left(\frac{P_{orb}}{\rm day}\right)^{-1/3} \left(\frac{M_2 \sin i}{M_J}\right)\ \rm ppm  \ , \\
\label{eq:aellip}
& \hspace{-10mm} A_{\rm ellip} =  \alpha_{\rm ellip} \frac{M_2 \sin i}{M_s} \left(\frac{R_s}{a}\right)^3 \sin i= 13\ \alpha_{\rm ellip} \sin i \left(\frac{R_s}{R_{\sun}}\right)^{3} \left(\frac{M_s}{M_{\sun}}\right)^{-2} \left(\frac{P_{orb}}{\rm day}\right)^{-2} \left(\frac{M_2 \sin i}{M_J}\right)\ \rm ppm  \ , \\
\label{eq:arefl}
& \hspace{-10mm} A_{\rm refl} =   \alpha_{\rm refl} 0.1 \left(\frac{R_s}{a}\right)^2 \sin i = 57\ \alpha_{\rm refl} \sin i  \left(\frac{M_s}{M_{\sun}}\right)^{-2/3} \left(\frac{P_{orb}}{\rm day}\right)^{-4/3} \left(\frac{R_2}{R_J}\right)^{2}\ \rm ppm \ , 
\end{align}
where $M_s$ and $R_s$ are the host star's (i.e. the primary star's) mass and radius, respectively, $M_2$ and $R_2$ are the secondary mass and radius, respectively, $P_{orb}$, $i$ and $a$ are the orbital period, inclination angle, and semimajor axis, respectively, $K_{RV}$ is the host star's radial velocity (RV) modulation semi amplitude, and $c$ is the speed of light. The three coefficients: $\alpha_{\rm beam}$, $\alpha_{\rm ellip}$, and $\alpha_{\rm refl}$ are each of order unity --- see \cite{faigler11} for more details.

The expressions above are approximations, where the primary assumptions are a circular orbit, negligible luminosity for the low-mass companion relative to the primary, and $M_2 \ll M_s$. 

\begin{figure}
\begin{center}
\includegraphics[scale=0.70]{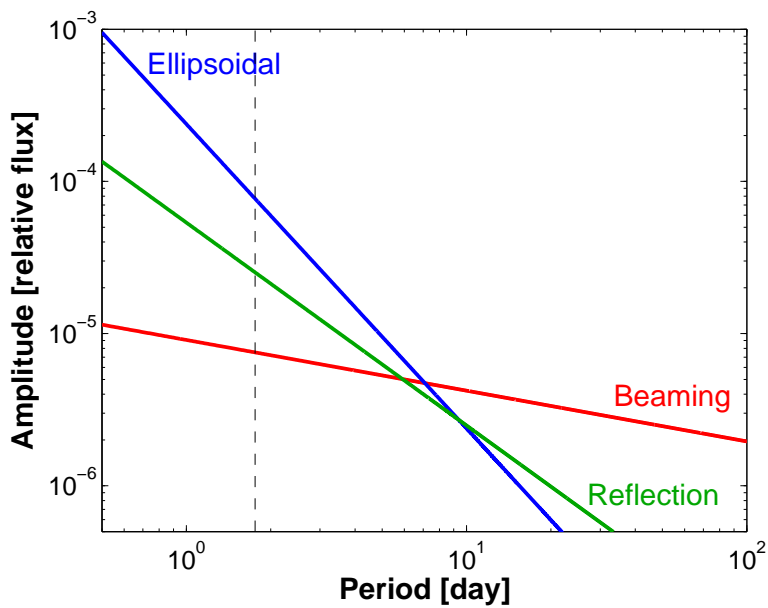}
\caption{\label{fig:effects} 
Expected amplitudes of the beaming (red), ellipsoidal (blue), and reflection (green) effects for a range of orbital periods, for a system similar to the KOI-13.01 star planet system. The plot is in log-log scale and is based on Eqs.~\ref{eq:abeam}--\ref{eq:arefl}. The dashed black line marks KOI-13.01 orbital period.
}
\end{center}
\end{figure}

Using the above three equations we show in \figr{effects} the expected amplitudes of the three effects for a star-planet system similar to KOI-13.01, and a range of orbital periods. The KOI-13.01 orbital period is marked by a vertical dashed line, where the expected amplitudes are an order of $10^{-5}$ to $10^{-4}$ in relative flux. We note that \figr{effects} is plotted while assuming the light from the target is not blended on \ik CCDs with light coming from nearby stars; such contamination will dilute the observed amplitudes.

\section{Data Analysis and Results}
\label{sec:dataanal}


We used here data from three \ik quarters, Q2, Q3, and Q5. Data from Q1 and Q4 were avoided as they were produced using insufficient aperture size to collect all the flux of this saturated star, and data from Q0 (and Q1) show increased noise level relative to other quarters. We verified that using all data through Q5 gave identical results, albeit with larger error bars by a factor of $\sim2$.

Before applying our analysis we applied two preprocessing steps. First, we ignored the in-eclipse points from all transits and occultations, removing 18.3 \% of the data.


The raw \ik data includes instrumental trends, temporal gaps and discontinuities (or ``jumps'') in the flux level. Correcting for these effects is important for detecting variability along the entire orbital phase, as opposed to the detection of eclipses whose typical duration is only a small fraction of the entire orbit. For that end we applied a second preprocessing step where we divided the \ik \lc into several continuous sections without gaps or jumps. The typical duration of each such section is 1--2 months. We fitted each section with a 5th degree polynomial and divided by the fit to get the corrected \lct, in relative flux. The fitting was done iteratively while ignoring 5 $\sigma$ outliers relative to the fit until no outliers were found. We checked that using a polynomial degree of 4 and 6 gave consistent results, within 0.5 $\sigma$. This step removed an additional 5.6~\% of the original three quarters data set, most of which from the start and end of the continuous sections.

\begin{figure}
\begin{center}
\includegraphics[scale=0.70]{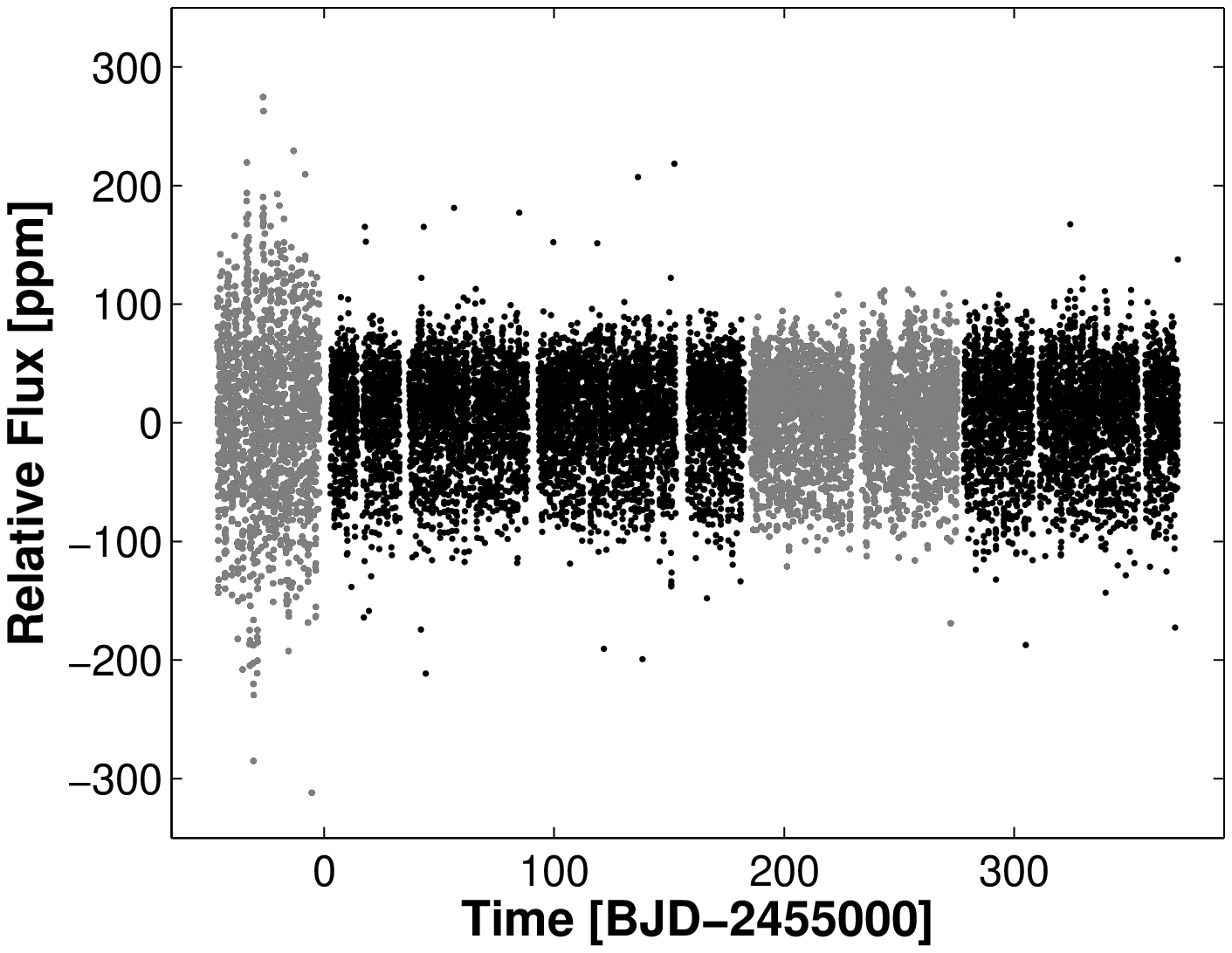}
\includegraphics[scale=0.70]{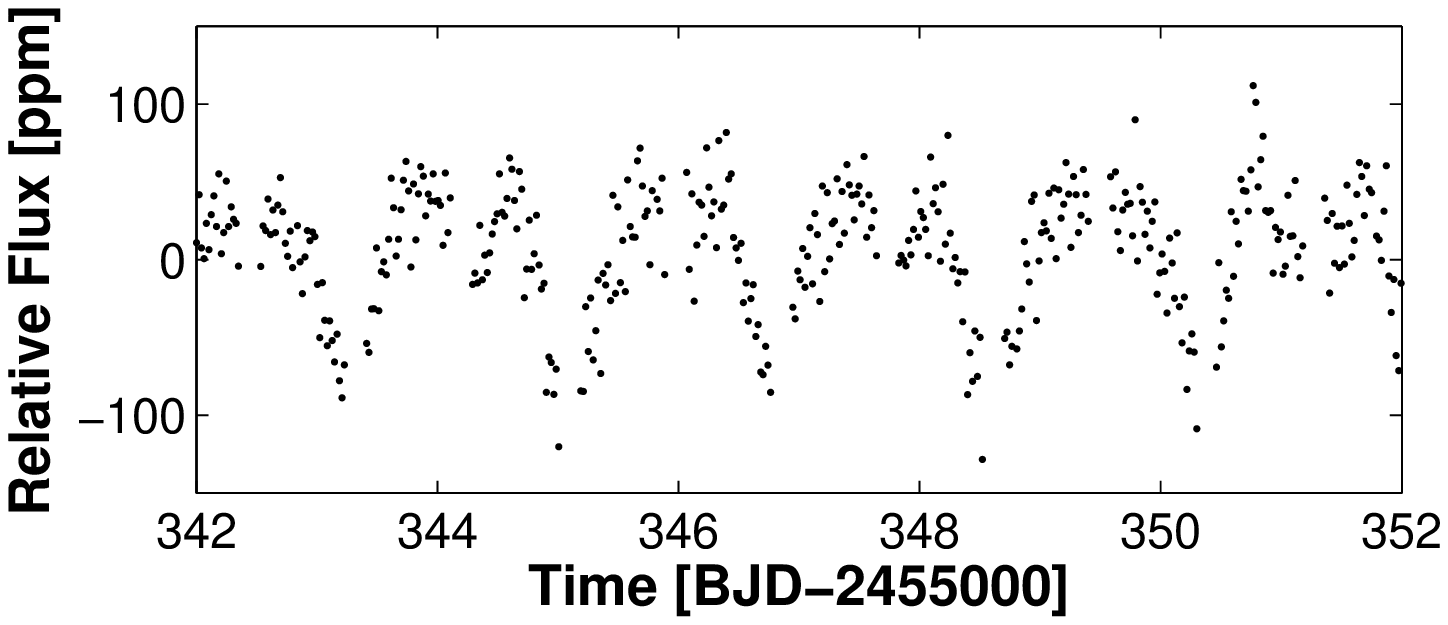}
\caption{\label{fig:lc} 
Top: KOI-13 \ik Long Cadence light curve, after applying the steps described in \secr{dataanal}, including removing the transit and occultation data. The data from Q0, Q1 and Q4 are shown in gray and were not used here. The data from Q2, Q3 and Q5 are shown in black. Bottom: Zoomed-in view of a 10 day light curve during Q5 (without in-eclipse measurements).
}
\end{center}
\end{figure}

The corrected \lc is shown in \figr{lc} top panel. Data from Q0, Q1 and Q4, that were not used in our analysis, are shown in gray. As can be seen in the figure the data from Q0 and Q1, containing the first 45 days of data have a larger scatter than the rest. The data we use in the following analysis, from Q2, Q3, and Q5, are shown in black. They contain 9,689 Long Cadence measurements, each with an effective exposure time of 29.4 minutes, spanning 368.4 days. \figr{lc} bottom planet shows a zoomed-in view of 10 days of data from Q5, where the orbital photometric variability can be visually identified.


To look for the orbital photometric signal, including the identification of  the orbital period and phase, and the measurement of the amplitudes of each of the photometric effects, we followed the 2-step BEER algorithm presented by \cite{faigler11}. 

\begin{figure}
\begin{center}
\includegraphics[scale=0.70]{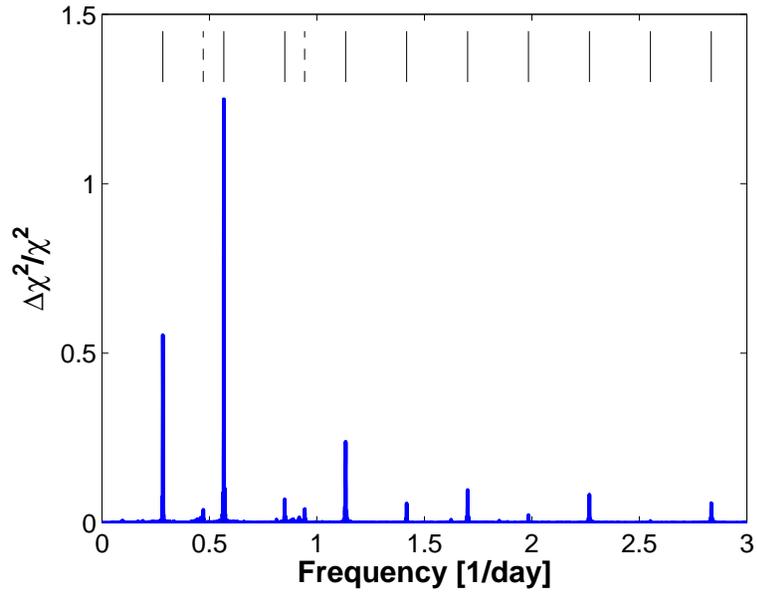}
\caption{\label{fig:per} 
The double harmonic periodogram (blue), where the strongest peak corresponds to the orbital period, of 1.76 days. The solid vertical lines (black) at the top of the figure mark frequencies which are a sub harmonic or harmonics of the strongest peak. The dashed vertical lines (black) mark unidentified frequencies where one is an harmonic of the other. See text for more details.
}
\end{center}
\end{figure}

{\it Step 1:} The goal of the first step is to detect the orbital period. To do so we constructed a periodogram where for each trial period we fitted, in a linear least squares manner, a double harmonic model \citep[e.g.,][]{shporer06,shporer10} including 5 parameters, or coefficients:
\begin{equation}
\label{eq:model}
f(t) = a_0 + a_{1c}\cos\left(\frac{2\pi}{P}t'\right) + a_{1s}\sin\left(\frac{2\pi}{P}t'\right) + a_{2c}\cos\left(\frac{2\pi}{P/2}t'\right) + a_{2s}\sin\left(\frac{2\pi}{P/2}t'\right)  , 
\end{equation}
where $f$ is relative flux, $P$ is a trial orbital period, and $t'=t-T_0$ where $t$ is time and $T_0$ is the time zero point determining the phase. $T_0$ is arbitrarily chosen in the first step and set to time of inferior conjunction\footnote{In a transiting system this is mid transit time.} in the second step (see below). For each trial period the statistics we used is based on the $\chi^2$ statistics and is taken to be:
\begin{equation}
\label{eq:stat}
\frac{\Delta \chi^2}{\chi^2} = \frac{\chi^2_{mean} - \chi^2}{\chi^2} , 
\end{equation}
where $\chi^2_{mean}$ is the $\chi^2$ for fitting a single parameter model, the measurements' mean. This method does not depend on an accurate knowledge of the measurement errors scaling. Our periodogram is shown in \figr{per}. The strongest peak corresponds to an orbital period of $P_{orb} = 1.7637 \pm 0.0013$ day. 
Peaks marked by a vertical solid black line are associated with $P_{orb}$, including harmonics and a sub-harmonic.


{\it Step 2:} Since the phase we used for the detection of the orbital period in the first step was arbitrary, the fitted coefficients do not necessarily have a one to one correspondence with the three physical effects described above. Therefore, in the second step we shifted $T_0$ to have phase zero positioned at inferior conjunction. This was done by using the fitted coefficients for the best fit period obtained in Step 1, and requiring that (I) $a_{2s}$, the only coefficient among the four that does not correspond to any physical effect, will vanish, and (II) $a_{2c}$ will be negative and $a_{1s}$ positive. The first requirement alone will still be ambiguous \citep[see][]{faigler11}. The derived transit time is $T_0$ = 24555138.7439 $\pm$ 0.0013 (BJD). The ephemeris derived here is consistent with the ephemeris of \cite{borucki11} for this object, and they are both listed in \tabr{ephem}. The higher precision of the \cite{borucki11} ephemeris is to be expected as it is derived from data covering the entire orbital phase, including transit and occultation.

After applying the shift we refitted the model using the period found in the first step, and the resulting fitted coefficients are listed in \tabr{coeff}. A quick examination of the coefficients shows their signs are each as expected, meaning negative for the reflection ($a_{1c}$) and ellipsoidal ($a_{2c}$) effects and positive for the beaming effect ($a_{1s}$).

We tried to estimate realistic errors on the fitted coefficients. The linear least squares method provides us ``formal" errors, but they do not account for possible remaining systematic trends in the \ik \lct.
To account for systematic trends we repeated our analysis while applying cyclic permutations to the residuals of our best fit model \citep[e.g.,][]{carter09}. Each of those permutations was generated by repeatedly shifting the residuals of every measurement and adding it to the model value of the subsequent point in the time series. The residual of the last measurement is added to the first point. Cyclic permutations keep the order of the residuals unchanged, preserving possible systematic trends, unlike random permutations. After analyzing all possible $10^4$ cyclic permutations we examined the distributions of the resulting coefficients and took half the difference between the 84.13 and 15.87 percentiles to be the 1 $\sigma$ uncertainty. The coefficient's distributions were highly symmetric about the median, and the latter were indistinguishable from the original fitted values. The errors derived this way were close to twice as large as the errors obtained directly from the linear least squares method. 

\begin{figure}
\begin{center}
\includegraphics[scale=0.7]{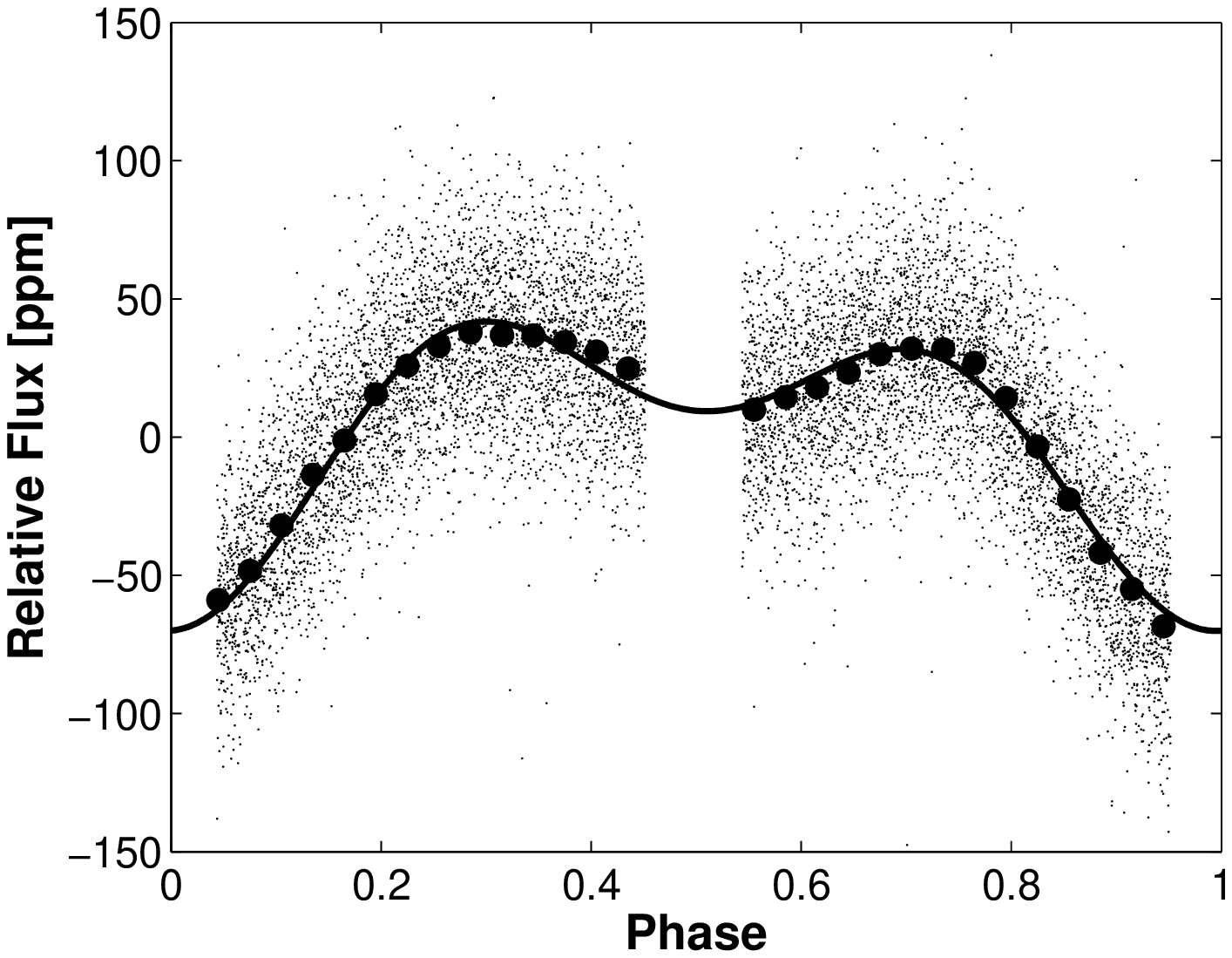}
\includegraphics[scale=0.7]{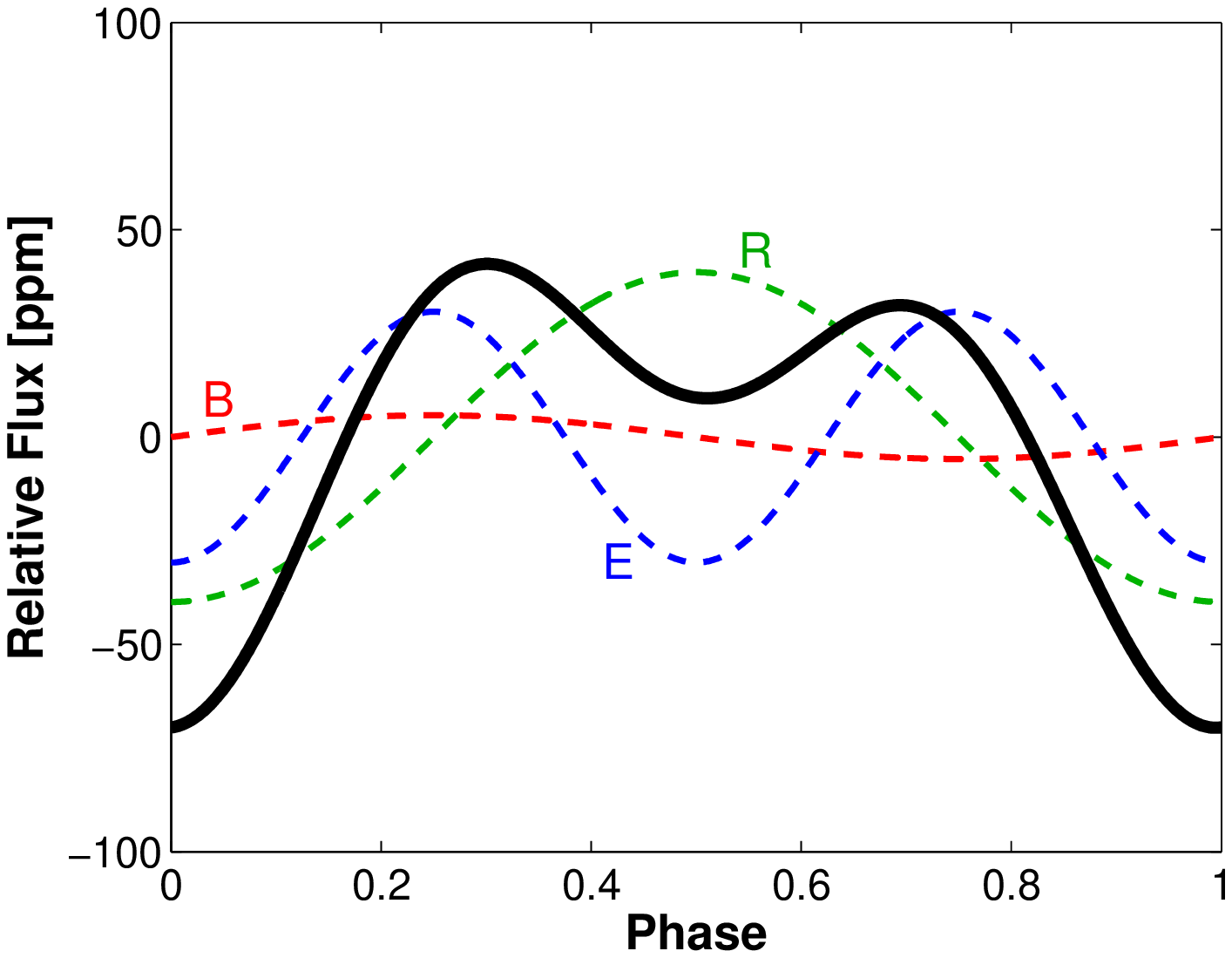}
\caption{\label{fig:foldlc} 
Top: phase folded light curve, using the ephemeris derived here. The gray dots are \ik Long Cadence data and the black circles are the binned light curve (error bars are smaller than the markers). The solid line represents the sinusoidal modulation constructed using the fitted coefficients. Bottom: we show the sinusoidal signals, in dashed lines, of each of the three effects plotted using the amplitudes found here (B=Beaming, E=Ellipsoidal, R=Reflection). The solid line (black) is the fitted double harmonic model, which is the sum of the three effects.
}
\end{center}
\end{figure}

Our phase folded \lct, based on the ephemeris derived here is presented in the top panel of \figr{foldlc}. That figure does not include the in-eclipse data, emphasizing that it was not part of the analysis. The sinusoidal modulation constructed from the coefficients fitted here is overplotted with a solid line and the filled circles represent the binned light curve. \figr{foldlc} bottom panel shows the individual sinusoidal effects plotted using the fitted coefficients, including the beaming (red), ellipsoidal (blue) and reflection (green) effects. The solid line (black) is the fitted model, the sum of the three modulations. The bottom panel shows that the overall double peak shape of the folded light curve results from the ellipsoidal effect, the large difference between the two minima results from the reflection effect and the small difference between the two maxima is the signature of the beaming effect. This panel shows clearly that the beaming effect amplitude is much smaller than the other two, which is expected according to \figr{effects}, although it is detected beyond a 10 $\sigma$ significance (see \tabr{coeff}).

\section{Discussion}
\label{sec:dis}


Dilution of the light from the target with light from a nearby star, or stars, will act to decrease the observed amplitudes of the three effects. Therefore, before using the fitted coefficients to extract physical information about the star-planet system one needs to account for possible dilution of light. For KOI-13 the dilution is significant since the target is the brighter member of a visual binary, with a separation of $\approx1$ arcsec. Since \ikt's pixels span 3.98 arcsec the light from the two stars is completely blended together. To translate our measured photometric amplitudes into the undiluted amplitudes, and in turn measure the systems physical parameters, we use the results of \citet[][see their Table 1]{szabo11}. We take their $V$ band magnitude difference, of 0.29 mag, to calculate the dilution, while adopting an error of 0.1 mag, since no error is given. We use the $V$ band magnitude difference as an approximated magnitude difference in the \ik band, since the two stars have similar $T_{eff}$, and they are located at the same distance and position in the sky. Therefore, the dilution factor, ${\cal D}$, by which the coefficients need to be multiplied in order to derive the undiluted amplitudes is $1.77 \pm 0.07$.


The amplitude of the beaming effect can be used to estimate the minimum mass of the secondary ($M_2 \sin i$). Rewriting \eqr{abeam} gives:
\begin{equation}
\label{eq:minmass}
M_2\sin i = \frac{0.37}{\alpha_{beam}} \left( \frac{M_s}{M_{\sun}}\right)^{2/3} \left( \frac{P_{orb}}{\rm day} \right)^{1/3} \left(\frac{A_{beam}}{\rm ppm}\right) \ M_J \ .
\end{equation}
We calculate the right hand side of \eqr{minmass} using $P_{orb}$ derived here, an undiluted beaming amplitude of $A_{beam}={\cal D} a_{1s} = 9.32 \pm  0.86$ ppm, and a host star mass of $M_s = 2.05\ M_{\sun}$ from \cite{szabo11}. We adopt a 10 \% uncertainty on the stellar mass since no error is given. The beaming coefficient is estimated to be $\alpha_{beam}=0.73 \pm 0.03$ assuming the host star radiates as a blackbody with $T_{eff}$= 8,510 $\pm\ 390$ K \citep{szabo11}, and by integrating over \ikt's transmission spectrum. The resulting minimum mass for the companion is $M_2\sin i=9.2 \pm 1.1\ M_J$. We note that a different estimate of  ${\cal D}$,  $M_s$ and $T_{eff}$ will result in a different $M_2\sin i$.

Since this system is transiting the above minimum mass is close to the true mass, and perhaps consistent with the latter within the error bars. This puts the companion's mass at the massive planet mass range.

As noted by \citet[][see also \citealt{kilic11}]{faigler11}, given sufficient information about the host star \eqr{aellip} shows that $A_{ellip}$ can be used to estimate another expression for the minimum mass, $M_2 \sin^2 i$, and combined with $M_2 \sin i$ estimated above the orbital inclination angle can be derived, which in turn allows to derive $M_2$. Meaning that analysis of orbital photometry signals can, in principle, result in determining the companion's true mass and not only the minimum mass. However, while \eqr{abeam}, describing the amplitude of the beaming effect is fairly accurate, \eqr{aellip}, describing the ellipsoidal effect amplitude is approximated \cite[e.g.,][]{morris85}. It was shown by \cite{pfahl08} that the amplitude of this effect for an early type star with no convective layer, like KOI-13, is difficult to describe analytically and it is likely far from the familiar expression used in \eqr{aellip} (although the shape of the effect can still be described as a cosine at the first harmonic of the orbital period). Therefore, we do not proceed here with using $A_{ellip}$ to derive $M_2$ and $i$. The lack of an adequate equation for $A_{ellip}$ does not allow us to use the ratio of the ellipsoidal and beaming amplitudes to check the likelihood of the model, as described by \cite{faigler11}.

\begin{figure}
\begin{center}
\includegraphics[scale=0.51]{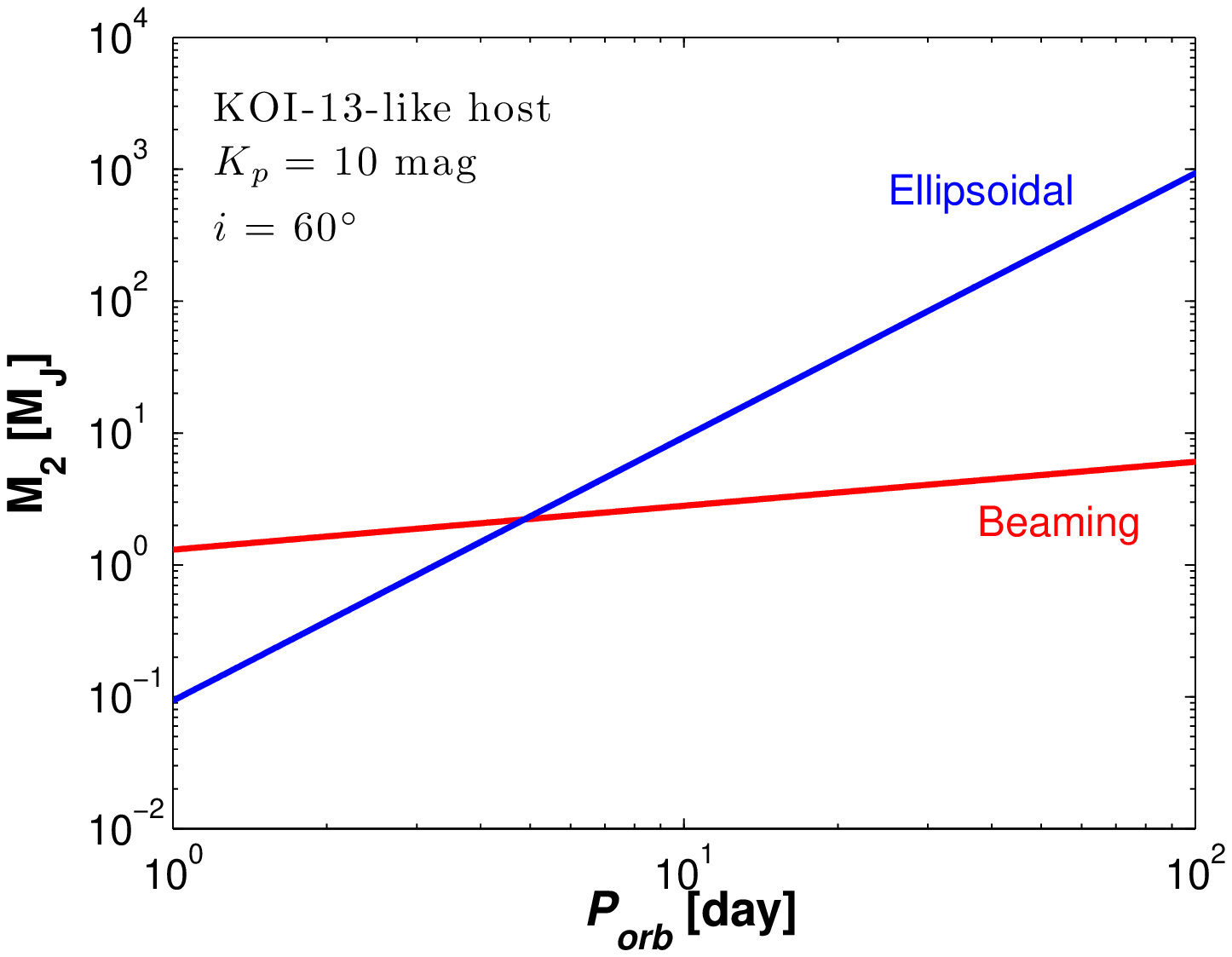}
\includegraphics[scale=0.51]{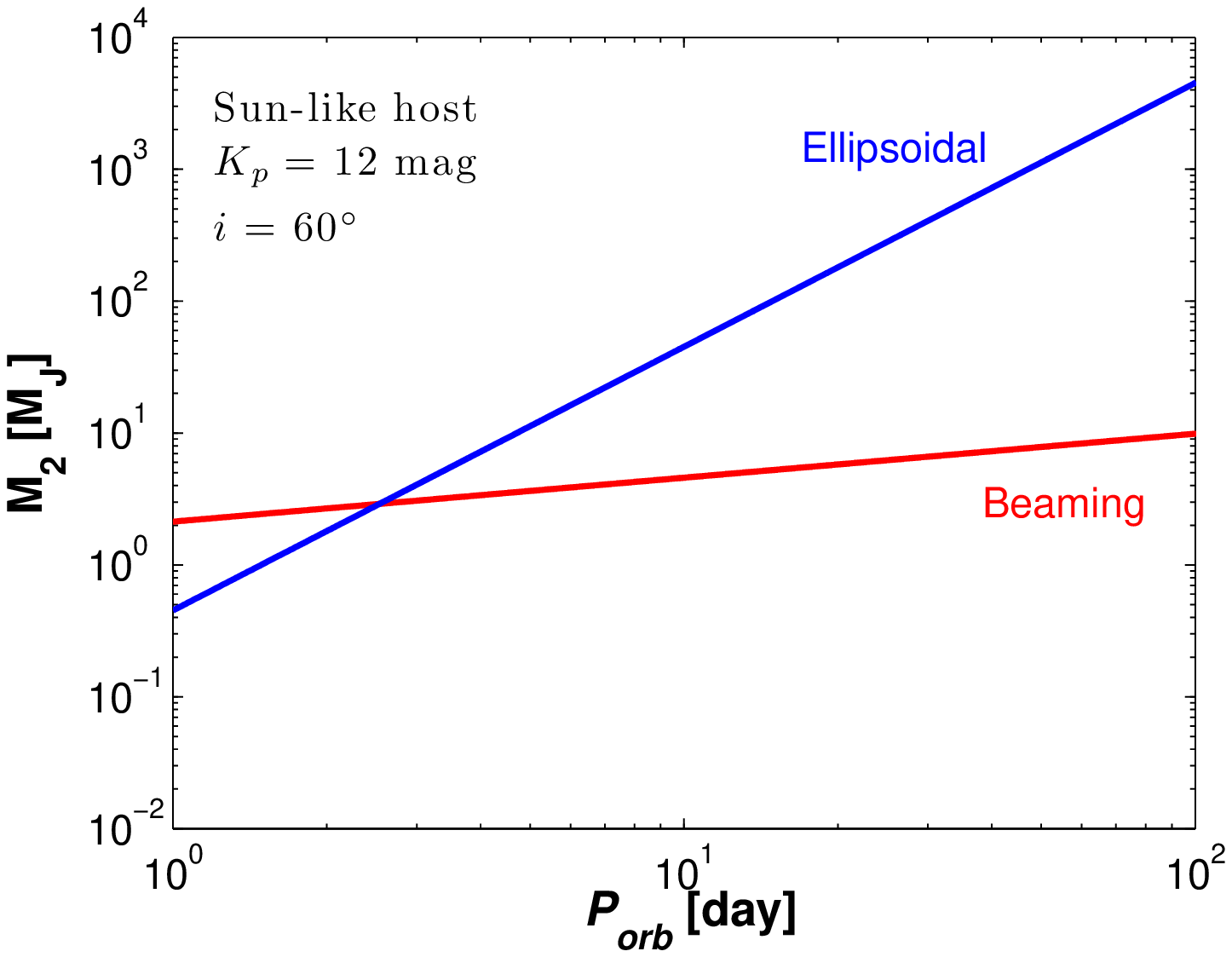}
\caption{\label{fig:sensitivity}
Minimum detectable secondary (star, brown dwarf or planet) mass as a function of orbital period for a 7 $\sigma$ detection, and assuming an inclination angle of 60 deg. The reflection effect can not be put on these diagrams, since its amplitude is sensitive to the companion's radius, which in turn is not sensitive to the companion's mass in the mass range of gas giants and brown dwarfs. Left: Assuming a KOI-13-like host (primary) star, and an amplitude of 2 ppm detected with \ik data from the entire 3.5 years mission lifetime. Right: Similar to the left panel, but for a Sun-like host star, at a brightness of $K_p=12$ mag and assuming 5 ppm amplitude.}
\end{center}
\end{figure}

The error bars derived for the fitted coefficients (see \tabr{coeff}) show that the KOI-13 light curve is sensitive to modulations down to 2 ppm which can be detected at  the 3--4 $\sigma$ level. Assuming white noise, the data from the entire 3.5 years \ik mission lifetime can be sensitive to 1 ppm modulations for a target similar to KOI-13 in brightness and stellar noise. It is interesting to use these sensitivities to calculate the companion's mass that can be identified for a range of orbital periods. To derive this predicted sensitivity of a search for  orbital photometry signals in \ik data we adopt a signal to noise threshold which is a factor of two larger than the above, bringing it to 7 $\sigma$. Although our error bars were already inflated by a factor of two relative to the formal error bars (see \secr{dataanal}), this additional factor of two is required in order to make the amount of false positives small enough considering the large number of light curves \citep[e.g.,][]{jenkins02}. Such sensitivity diagrams are shown in \figr{sensitivity}. The left panel is for a KOI-13-like host, and the right panel is for a Sun-like host at $K_p$=12 mag. Only the beaming and ellipsoidal effects are  plotted on those diagrams, since they are sensitive to the companion's mass. The reflection effect is sensitive to the companion's radius (see \eqr{arefl}), which is only weakly dependent on mass at the mass range of gas giant planets and brown dwarfs. Both diagrams assume no dilution and an arbitrarily chosen orbital inclination angle of $i=60$ deg. The lines in both panels correspond to the expected sensitivity once data from the entire mission lifetime is obtained, taken to be 2 ppm in the left panel and 5 ppm in the right panel, detected at the 7 $\sigma$ level. 

A close inspection of \figr{sensitivity} shows that orbital photometry is sensitive to short-period low-mass companions down to the few Jupiter mass level. Therefore, it can be used to detect low-mass stellar companions, brown dwarfs and massive planets. This mass range completely overlaps the so called brown dwarf desert --- a paucity in the mass function of companions to Sun-like stars \citep[e.g.,][]{grether06, sahlmann11, evans11}. Occupants of the brown dwarf desert are rare, but orbital photometry can detect most if not all short-period brown dwarfs orbiting bright stars in the \ik field, thereby characterizing the high mass end of the planetary mass distribution and the low mass end of stellar companions mass distribution.

The assumption of white noise above and the sensitivities presented in \figr{sensitivity} could be somewhat optimistic. Later type stars, colder than KOI-13, constitute the majority of \ik targets and are expected to show activity at a level that may exceed the \ik precision. This will degrade the sensitivity of the BEER algorithm to the orbital photometric signal. Stellar oscillations also represent a potential problem for the BEER approach, although unless the oscillating frequency is closely related to the orbital frequency it is possible they can both be resolved. Both stellar activity and oscillations could in principle mimic an orbital photometric signal, resulting in a false BEER detection. Therefore, BEER detections should be examined carefully, specifically that the coefficients' sign and ratio are as expected (for more details see \citealt{faigler11}). In addition, many of the \ik targets are blended with other nearby stars on \ik CCDs, as is KOI-13 itself. Blending will decrease the overall amplitude of the orbital signal but otherwise will not affect its shape, resulting in underestimating the companion's mass. All the above emphasize the need for RV follow-up to confirm BEER detections and measure the companion's mass.

On a more positive note, the errors listed in \tabr{coeff} were obtained from an incomplete light curve, without the transit and occultation phases which cover $\approx 18$ \% of KOI-13.01 orbit. Therefore, the results obtained here and the sensitivities shown in \figr{sensitivity} should be somewhat better for a non-transiting system similar to KOI-13.01, where the entire light curve is used, unless the system's orientation is close to face-on. 

Analysis of orbital photometry can, in principle, be used to look for additional signals, induced by additional objects in the system, including a possible second planet which is not necessarily transiting. This possibility is intriguing in light of the large number of candidate multi transiting planet systems found by \ik \citep{lissauer11}, and the fact that about one sixth of the candidate transiting planetary systems contain more than a single transiting planet \citep{borucki11}. \cite{latham11} have shown that massive planets, like KOI-13.01, are less likely to reside in multi transiting planet systems, although it is still possible that additional planets occupy non-transiting orbits.

Our periodogram presented in \figr{per} shows two small peaks, marked by dashed vertical lines, that are not associated with the 1.76 day orbital period. The strongest among the two corresponds to a period of $1.0600 \pm 0.0010$ days, and the smaller peak is at the first sub-harmonic. We verified this secondary photometric signal, whose semi amplitude is 12 ppm, exists in each of the three \ik quarters included in our analysis.

We investigated this further by subtracting the 1.76 day double harmonic model and applying the same period analysis as described above to the residuals. The derived fitted coefficients were $a^{(2)}_{1c} = 2.3 \pm 1.0$ ppm, $a^{(2)}_{1s} = 11.8 \pm 1.0$ ppm, and $a^{(2)}_{2c} = -3.1 \pm 0.5$ ppm, where the upper index is used for marking a secondary photometric signal. The sign of $a^{(2)}_{1c}$ and the relative size of the coefficients makes it unlikely this signal originates from a second planet orbiting at 1.06 days. 

The ratio between the 1.06 day period and the orbital period is $0.60100 \pm 0.00072$, which is close to an integer numbers ratio of 3:5. Assuming this is not a coincidence, it raises the possibility that the 1.06 day signal is a stellar pulsation triggered by the planetary companion. A similar scenario was suggested for WASP-33 \citep{collier10}, a planetary system that resembles KOI-13 as it includes a short period (1.22 days) transiting planet orbiting a pulsating A-type star whose spin axis is misaligned with the planetary orbital axis \citep{collier10, barnes11}. An extreme example of pulsations induced by a binary companion is the KOI-54 binary system \citep{welsh11}, composed of two similar A-type stars on a highly eccentric orbit. Still, this scenario does not explain why we see a periodicity specifically at a relation of 3:5 with the orbit and not another integer relation.

We conclude that the true nature of this secondary photometric signal is currently not clear, although unlikely to be due to a second planet. In addition, we note that since the target is blended with a nearby star on \ik CCDs, this secondary photometric signal could originate from the nearby star. A more detailed study of this signal, once more \ik data becomes available, will be important for better understanding the possible false alarm scenarios of the application of the BEER approach to \ik data, which have not been thoroughly studied yet.


Finally, we note that a close inspection of the phase folded light curve in \figr{foldlc} top panel shows there are small deviations of the binned light curve from our double harmonic model. That model is simplistic, meant to be used only as a detection tool, and not as a detailed physical model, especially for data of such high quality. Therefore, the presence of these deviations should not be alarming, but rather be viewed as an opportunity for a more detailed study of bright systems using \ik data.

\section{Summary}
\label{sec:sum}

We have presented here a test case for analyzing orbital photometry using the BEER approach \citep{faigler11}. We have shown that photometric measurements taken out of eclipse (transit and occultation) can be used to detect the presence of KOI-13.01, measure its orbital ephemeris and estimate its minimum mass. Therefore, similar {\it non-transiting} systems can be detected using \ik data.

Moreover, we used the error bars derived for the fitted coefficients and a 7 $\sigma$ detection threshold to estimate the sensitivity of \ik data for low-mass non-transiting companions. Assuming white noise we showed that data accumulated during the entire 3.5 years \ik mission lifetime will allow detection of low-mass short-period companions down to the few Jupiter mass level, for bright stars. In case the \ik mission will be extended, this sensitivity will improve, allowing detection of lower mass companions, orbiting fainter stars at longer periods.

Spectra of hot stars, of early spectral type like KOI-13, are not rich with deep and narrow absorption lines which are used for measuring high precision RVs. As a result those stars are not regularly monitored by RV surveys, leading in turn to highly limited knowledge about their low-mass companion population. Analysis of the orbital photometry signal presents an opportunity to extend our knowledge about low-mass companions to early type host stars, especially considering they show a decreased amount of stellar activity. For example, orbital photometry can be used to test the suggestion of \cite{bouchy11}, that massive planets and brown dwarfs are more frequent around F-type stars than G-type stars, and see if this trend continues into A-type stars. RV confirmation of such companions, detected through orbital photometry, will probably be difficult or even impossible (For an RV confirmation of a white dwarf companion detected through orbital photometry see \citealt{ehrenreich11}). However, a detection of a large sample along with an improved understanding of the false positive scenarios will allow a statistical analysis and a comparison to the characteristics of the companion population orbiting Sun-like stars.

The analysis done here uses only out of eclipse data. Several authors have already shown that \ikt's continuous and high precision photometry taken {\it in-eclipse} can be used for additional characterization of the companion's orbit \citep{barnes09, barnes11, shporer11, groot11, szabo11}. These authors show that small asymmetries in the eclipse light curve can be used for measuring the sky-projected spin-orbit angle --- angle between the star's rotation axis and the companion's orbital angular momentum axis. Thereby showing the high scientific potential of using photometry, out and in eclipse, to study binary systems, those with low-mass companions in particular.

\acknowledgments

\ik was competitively selected as the tenth Discovery mission. Funding for this mission is provided by NASAÕs Science Mission Directorate.
We warmly thank Michael Endl for useful comments. 
A.S. acknowledges support from NASA Grant Number NNX10AG02A.


{\it Facilities:} \facility{The \ik Mission}

\begin{deluxetable}{lcc}
\tablecaption{\label{tab:ephem} KOI-13.01 ephemeris} 
\tablewidth{0pt}
\tablehead{\colhead{Parameter} & \colhead{This work} & \colhead{\cite{borucki11}} }
\startdata
\hline
Orbital period, $P_{orb}$ (days) 		& \ \ \ \ \ \ \ \ \  1.7637 $\pm$ 0.0013 & \ \ \ \ \ \ \ \ \  1.7635892 $\pm$ 0.0000014 \\
Inferior conjunction time, $T_0$ (BJD)	&	2455138.7439 $\pm$ 0.0013 & 2454953.56498 $\pm$ 0.00012 \\ 
\hline
\enddata
\end{deluxetable}

\begin{deluxetable}{ccc}
\tablecaption{\label{tab:coeff} Fitted coefficients} 
\tablewidth{0pt}
\tablehead{\colhead{Coefficient} & \colhead{Effect} & \colhead{Value} \\
 &  & \colhead{[ppm]} }
\startdata
\hline
$a_{1c}$ 	&	Reflection		&	-39.78 $\pm$ 0.52\\
$a_{1s}$ 	&	Beaming		&	\ \  5.28  $\pm$  0.44 \\
$a_{2c}$ 	&	Ellipsoidal	&	-30.25 $\pm$ 0.62\\
$a_{2s}$ 	&	---			&  \  \ \ \ 0.0	$\pm$ 0.48 \\
\hline
\enddata
\end{deluxetable}

\end{document}